\pdfoutput=1
\RequirePackage{fix-cm}
\documentclass[british]{article}

\usepackage{spconf} % Conference template

\usepackage[utf8]{inputenc}
\usepackage[T1]{fontenc}
\usepackage[british]{babel}

\usepackage{pifont}

\usepackage{graphicx}
\usepackage{xcolor}
\graphicspath{{./figures/}}
\DeclareGraphicsExtensions{.pdf,.png,.jpeg,.jpg}

\usepackage{amsmath} % cmex10 ensures that only type 1 fonts are used at all font sizes
\usepackage{amssymb}
\usepackage{bm} % Bold maths
\usepackage{esint} % More integrals for computer modern math fonts
\usepackage{commath} % Provides \dif upright d
\usepackage{nicefrac} % Provides \nicefrac

\usepackage{subcaption} % Recommended, e.g., with the latest ACM template

\usepackage{booktabs} % Better-looking, more formal tables
\usepackage{array} % Patches and improves the array and tabular environments

\usepackage{multirow} % Enables table cells to span more than one row
\usepackage{rotating} % Enables rotating text in tables

\usepackage{enumitem} % Better control of lists and their spacing
\setlist{nolistsep} % Removes vertical spacing in lists

\usepackage[unicode=true,pdfusetitle,%
 pdfauthor={Shivam Mehta, Ruibo Tu, Simon Alexanderson, Jonas Beskow, Éva Székely, Gustav Eje Henter},%
 pdfkeywords={text-to-speech, co-speech gestures, speech-to-gesture, integrated speech and gesture synthesis, ODE models},%
 bookmarks=true,bookmarksnumbered=true,bookmarksopen=false,%
 breaklinks=true,pdfborder={0 0 0},backref=false,colorlinks=true,citecolor=blue]{hyperref}

\usepackage[capitalise]{cleveref} % Use \cref to automatically include, e.g., "Sec." in \ref (must be loaded after hyperref)
\Crefname{section}{Section}{Sections}
\crefname{section}{Sec.}{Secs.}
\Crefname{figure}{Figure}{Figures}
\crefname{figure}{Fig.}{Figs.}
\Crefname{table}{Table}{Tables}
\crefname{table}{Table}{Tabs.}

\renewcommand{\mid}{\,\ifnum\currentgrouptype=16 \middle\fi|\,}

\newcommand{\customurl}[1]{\url{#1}} % Monospaced URLs

\newcommand{\webpageurl}{https://shivammehta25.github.io/Match-TTSG/}
\newcommand{\webpageurltext}{shivammehta25.github.io/Match-TTSG/}

\ninept % Use minimum tolerated font size
\let\oldmarginpar\marginpar
\renewcommand\marginpar[1]{\-\oldmarginpar[\raggedleft\footnotesize #1]%
{\raggedright\footnotesize #1}}

\title{Unified speech and gesture synthesis using flow matching}

\name{Shivam Mehta, Ruibo Tu, Simon Alexanderson, Jonas Beskow, Éva Székely, Gustav Eje Henter\thanks{This work was partially supported by the Wallenberg AI, Autonomous Systems and Software Program (WASP) funded by the Knut and Alice Wallenberg Foundation, by the Digital Futures project ``Artificial Actors: Directable digital humans based on psychological models of relational reasoning'', and by the Industrial Strategic Technology Development Program (grant no.\ 20023495) funded by MOTIE, Korea.}}
\address{Division of Speech, Music and Hearing, KTH Royal Institute of Technology, Stockholm, Sweden}
\begin{document}
\maketitle
\begin{abstract}
%The abstract should contain about 100 to 150 words
%The online interface says "Limit the abstract text to a maximum of 200 words."
As text-to-speech technologies achieve remarkable naturalness in read-aloud tasks, there is growing interest in multimodal synthesis of verbal and non-verbal communicative behaviour, such as spontaneous speech and associated body gestures. This paper presents a novel, unified architecture for jointly synthesising speech acoustics and skeleton-based 3D gesture motion from text, trained using optimal-transport conditional flow matching (OT-CFM). The proposed architecture is simpler than the previous state of the art, has a smaller memory footprint, and can capture the joint distribution of speech and gestures, generating both modalities together in one single process. The new training regime, meanwhile, enables better synthesis quality in much fewer steps (network evaluations) than before. Uni- and multimodal subjective tests demonstrate improved speech naturalness, gesture human-likeness, and cross-modal appropriateness compared to existing benchmarks.
%in joint speech and gesture synthesis.
\end{abstract}
\begin{keywords}%
Text-to-speech, co-speech gestures, speech-to-gesture, integrated speech and gesture synthesis, ODE models
\end{keywords}
\section{Introduction}
\label{sec:intro}

Human communicative behaviour builds on a rich set of carefully orchestrated verbal and non-verbal components.
Speech, facial expressions, and gestures are fundamentally intertwined, born out of a shared mental representation of communicative intent \cite{mcneill2008gesture} and coloured by context such as emotions and the communicative situation at hand.
Gestures can reinforce, complement, or entirely replace spoken words \cite{kendon1988gestures,mcneill2008gesture}, affecting the interpretation of utterances.
The presence of gestures improves to understanding and recall
%\cite{rogers1978contribution,riseborough1981physiographic,hostetter2011gestures},
with positive effects that carry over to human-computer interaction \cite{nyatsanga2023comprehensive}.

It is well known that there is highly precise synchronisation, e.g., between individual gestures, facial expressions, and prosodic inflections \cite{wagner2014gesture}.
At the same time, non-verbal behaviours are highly variable, even optional, which makes them impossible to predict deterministically, and the variation itself appears to be a fundamental property of these behaviours.
Generating naturalistic, meaningful, and coherent conversational behaviours with a spontaneous quality
%(i.e., not scripted, repetitive, or robot-like),
is crucial to many applications, e.g., in extended reality, telepresence, computer games,
%pedagogical agents,
virtual assistants, and social robots.

%Speech- and gesture synthesis have to a large extent been treated as separate problems by disjunct research communities, and for different communicative contexts; whilst speech synthesis for the most part has focussed on modelling reading-aloud, gestures studies generally focus on spontaneous production and interaction, which is where the majority of gesturing occurs.

Today, unimodal synthesis of speech audio can obtain a quality virtually indistinguishable from human production of isolated utterances \cite{shen2018natural, mehta2023overflow}, and reach striking human-likeness even for spontaneous speech \cite{szekely2020breathing}.
The same is increasingly true for 3D gesture motion \cite{alexanderson2023listen, ao2023gesturediffuclip}.
However, making the two modalities appropriate for each other -- important for consistent and effective communication -- is largely an unsolved problem \cite{yoon2022genea, mehta2023diff, kucherenko2023genea}.

%Taken together, the above findings suggest that
In summary, speech and gesture (being non-deterministic) should be synthesised using powerful probabilistic models, and also modelled together, as they stem from a common representation.
This motivates the main contributions of this paper, which are:
\begin{enumerate}
\item Proposing a truly unified architecture for speech and gesture generation that describes (and samples from) the joint probability distribution of speech and 3D gestures.
This is simpler than previous approaches, and significantly better for generating speech and gestures that are in agreement with each other.
\item Training the new architecture using conditional flow matching (OT-CFM) \cite{lipman2023flow}.
This learns ODE-based representations of probability distributions, similar to probability flow ODEs from diffusion models \cite{song2021score}, but constructed so as to make the differential equations easy to solve numerically.
This leads to substantial improvements in speed and realism.
\end{enumerate}
Experiments demonstrate that our two contributions together yield a 15-fold speedup over the previous state-of-the-art method, Diff-TTSG \cite{mehta2023diff}, and use less memory, whilst simultaneously producing significant improvements on all three subjective performance measures assessed in our user studies.
%We evaluate our approach both objectively, in terms of synthesis speed and ASR word error rate, and subjectively with uni-modal and multimodal user studies, disentangling synthesis quality in different modalities from their appropriateness for each other, and find that the proposed method represents significant improvements over the prior state of the art, Diff-TTSG \cite{mehta2023diff}, across all metrics.
We call our new approach \emph{Match-TTSG} due to its use of flow matching.
%The results demonstrate improved speech and gesture quality and better alignment between the two modalities (increased gesture appropriateness for the speech), compared to the previous state of the art.
%in integrated speech and gesture.
%Example videos are enclosed as supplementary material and will be made available online upon paper acceptance, along with code.
% Example videos, to be made public on acceptance, are in the supplement and at anonymous URL \url{bit.ly/diff-ttsg}.
%, and will be made public upon acceptance.
%, together with code.
%And demonstrate
For video examples and code, please see \href{\webpageurl}{\webpageurltext}.

\section{Background}
\label{sec:background}

\begin{figure*}[t]
  \centering
  \includegraphics[width=0.9\linewidth]{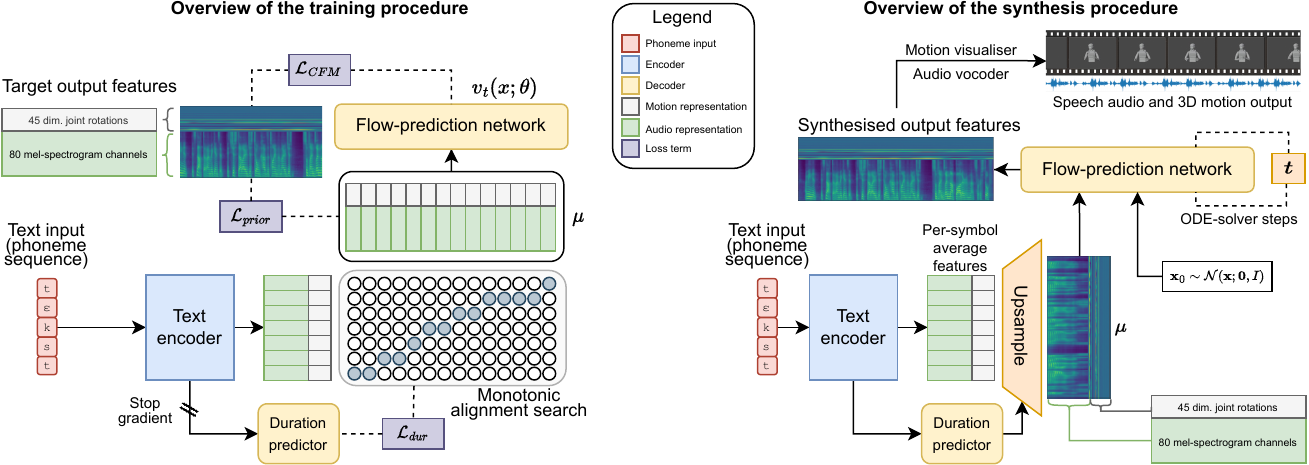}
  %\begin{subfigure}{\columnwidth}
  %  \includegraphics[width=0.85\linewidth]{training_compact.pdf}
  %  \caption{Overview of the training procedure. Dashed lines indicate loss terms.}
  %  \label{fig:training overview compact}
  %\end{subfigure}
  %\hfill
  %\begin{subfigure}{\columnwidth}
  %  \includegraphics[width=0.85\linewidth]{synthesis_compact.pdf}
  %  \caption{Overview of the synthesis procedure.}
  %  \label{fig:synthesis overview compact}
  %\end{subfigure}
  \caption{Schematic overview of Match-TTSG training and synthesis. Still frames of the avatar used to visualise motion can be seen top right.}
  \label{fig:architectural overview compact}
  \vspace{-\baselineskip}
\end{figure*}

%Currently, some of the highest-quality TTS systems utilise deep diffusion models \cite{popov2021grad, betker2023better}, discrete-time normalising flows \cite{kim2021vits,mehta2023overflow}, and \emph{conditional flow matching} (CFM) \cite{le2023voicebox}.
\subsection{Prior work}
\label{ssec:prior}
Deep generative methods including diffusion probabilistic models (DPMs) are currently the state of the art in text-to-speech (TTS) \cite{popov2021grad,kim2021vits,mehta2023overflow} and speech-to-gesture tasks \cite{alexanderson2020style,alexanderson2023listen,ao2023gesturediffuclip}.
However, relatively few models have been proposed to generate 3D motion together with speech. 
%-- in contrast to audiovisual speech synthesis of audio and 2D images/video of talking faces, which is more established \cite{mattheyses2015audiovisual}. 
A handful of recent works have considered synthesising talking heads \cite{yu2019durian} or facial landmarks \cite{mitsui2023uniflg} in 3D from text simultaneously with speech acoustics.
Here, we instead focus on the joint synthesis of speech acoustics and skeleton-driven 3D-character body motion from text.
This can be more challenging than generating facial motion (specifically lips), since lip motion is relatively predictable from speech acoustics, whereas body gestures are not.
%However, relatively few models have been proposed to generate 3D motion together with speech -- while audiovisual speech synthesis of talking faces has a longer history (see \cite{mattheyses2015audiovisual} for and overview), with \cite{yu2019durian,mitsui2023uniflg} as recent examples. The focus of this article, however, is on the joint synthesis of speech acoustics and skeleton-driven 3D-character body motion from text. This can be considered a more challenging problem than generating facial motion (specifically lips), since lip motion is relatively closely determined by speech acoustics, whereas body gestures are not.

%Only a limited amount of prior work exists when it comes to
For integrated generation of speech and gesture, only limited prior work exists.
Instead, speech- and gesture synthesis have largely been treated as separate problems by separate research communities, considering different communicative contexts:
text-to-speech has mostly focused on read-aloud speech, whilst the majority of co-speech gesturing occurs (and its research) in spontaneous production and interaction.
%Because of the mathematical similarity between the acoustic features produced by a TTS acoustic model and 
%The typical approach to multimodal synthesis has been to start from a TTS model, since these models can solve the alignment problem between input symbols (phones) and output features. %Salem et al \cite{salem2010towards} integrated speech and gesture production via a tight coupling between a TTS system and the perceptuo-motor system of a robot. 
%, wheras speech-to-gesture models can generally assume that audio and gestures are aligned in time.
An early proposal to unite the two was \cite{salem2010towards}, with the first coherent integration of deep-learning-based speech and gesture models being \cite{alexanderson2020generating}.
They trained separate text-to-speech and speech-to-gesture models on the same multimodal recordings, and connected the systems into a two-stage synthesis pipeline to produce speech and gestures using the same speaker identity and spontaneous style.

Wang et al.\ \cite{wang2021integrated} went a step further, and adapted both Glow-TTS \cite{kim2020glow} and Tacotron 2 \cite{shen2018natural} for multimodal speech-and-gesture synthesis.
The latter approach was found to produce the best results, but required numerous accommodations to work, including the use of pre-trained unimodal models and multi-stage fine-tuning with parts of the network frozen.
Most recently, Diff-TTSG \cite{mehta2023diff} used probabilistic denoising diffusion models based on Grad-TTS \cite{popov2021grad}.
%in an architecture that allows for
They were able to train high-quality speech-and-gesture synthesis simultaneously from scratch, but used separate diffusion pathways for the two modalities at synthesis time.
%Mathematically, the consequence of this design is that the model assumes conditional independence between the two modalities:
This is mathematically equivalent to assuming conditional independence between the two modalities:
\begin{align*}
P(\text{acoustics},\,\text{motion} \mid \text{text})
& = P(\text{acoustics} \mid \text{text}) \cdot P(\text{motion} \mid \text{text})
\text{.}
\end{align*}
Diff-TTSG synthesis thus combines independent samples from two different marginal distributions.
This can lead to reduced cross-modal appropriateness, e.g., acoustic prominence may randomly be assigned to one word but gestural prominence assigned to another.
Furthermore, the motion diffusion pathway used 500 ODE-solver steps to obtain good output quality, making the model slow to run.

We here present the first truly unified approach for integrated speech and gesture generation that can synthesise high-quality output from the joint distribution of the two modalities.
In addition, it requires less memory to train and is much faster to run.
%is highly efficient both for training and synthesis.

\subsection{Conditional flow matching}
\label{ssec:cfm}
%Currently, some of the highest-quality TTS systems utilise deep diffusion models \cite{popov2021grad, betker2023better}, discrete-time normalising flows \cite{kim2021vits,mehta2023overflow}, and \emph{conditional flow matching} (CFM) \cite{le2023voicebox}.
%Among them, CFM-based models achieve the state of the art results, but they are still without sufficient exploration for TTS, motion generation, and the combination of them. 
%To our knowledge, the only CFM-based speech synthesis model is Voicebox (VB) \cite{le2023voicebox}, and there is no model for either motion synthesis or the combination of speech and motion synthesis.

Deep generative models have recently been demonstrating strong results by learning and sampling from distributions represented as initial-value problems with ordinary differential equations (ODEs).
Examples include continuous-time normalising flows (CNFs) \cite{chen2018neural} and probability flow ODEs derived from DPMs \cite{song2021score}.
Lipman et al.\ \cite{lipman2023flow} proposed \emph{conditional flow matching} (CFM), a new, general framework for learning ODEs that represent distributions.
They also proposed \emph{optimal transport CFM} (OT-CFM), a particular version of their framework that is very similar to the concurrent idea of \emph{rectified flow} \cite{liu2023flow}.
This leads to particularly simple -- nearly constant -- ODE vector fields for synthesis.
Such vector fields are not only easier for neural nets to learn than those originating from diffusion models,
%(and can be trained without using adjoint sensitivity methods, unlike \cite{chen2018neural}),
but
%the linear nature also leads to largely constant vector fields
since the fields do not change much, the ODE (which is used when sampling from the learnt distribution) can be accurately solved with a numerical ODE solver using very few discretisation steps, and correspondingly fewer neural-network evaluations.
%Since all solution steps are sequential and each requires evaluating a neural network, this property
This enables a very substantial synthesis-time speedup over DPM-based approaches.
%This leads to much lower number of function evaluations (a.k.a.\ synthesis steps) than deep diffusion models \cite{song2021score} with a similar synthesis quality.

OT-CFM was first applied to
%text-conditional
speech synthesis in \cite{le2023voicebox}.
%An application of OT-CFM to 3D motion synthesis.
%However, to the best of our knowledge, our
We believe our article is the first example of CFM applied to 3D motion synthesis as well as multimodal synthesis.
A preprint applying CFM to unimodal 3D motion \cite{hu2023motion} appeared whilst our paper was in review.
%A preprint \cite{hu2023motion} concurrently applying CFM to unimodal motion appeared whilst this paper was in review.
%continuous normalising flows (CNFs) \cite{chen2018neural}, which synthesise data based on solving initial value problems of ordinary differential equations (ODEs).

%They first constructed conditional vector fields that can be used for generating data following the data distribution, and applied the regression loss (CFM) to learning the constructed conditional vector fields. They can synthesise data by solving the initial value problems defined by the learned vector field. 
%Due the simplicity of the regression task, CFM is much more efficient than learning score functions \cite{song2021score} and than using the adjoint sensitivity methods \cite{chen2018neural}. Moreover, unlike deep diffusion models \cite{song2021score} and continuous normalising flow in \cite{chen2018neural}, conditional vector fields in CFM-based models \cite{lipman2023flow} can be constructed in a flexible way; hence, one can utilise a nearly constant conditional vector field inspired by optimal transport as in OT-CFM \cite{lipman2023flow}. 
%Such simple vector field means that the initial value problem can be accurately solved in few discretisation steps with an ODE numerical solver, which leads to much lower number of function evaluations (or synthesis steps) than deep diffusion models \cite{song2021score} with a similar synthesis quality.

\section{Method}
\label{sec:method}
Both speech and gesture motion can be mathematically represented as vector-valued time series.
For speech, these may be speech acoustics $\boldsymbol{a}_{1:T}$ (e.g., mel spectra), whilst motion $\boldsymbol{m}_{1:T}$ is described by a sequence of poses, here vectors containing rotations for joints on a 3D skeleton.
If speech and motion features are sampled at the same frame rate, they can be concatenated as $[\boldsymbol{a}^\intercal,\boldsymbol{m}^\intercal]^\intercal_{1:T}$ and may be synthesised by a single process (e.g., ODE) in a unified manner.
Despite that, previous efforts have not been fully integrated in this way.

Although \cite{wang2021integrated} investigated Glow-TTS \cite{kim2020glow} for synthesising concatenated audio and motion features, results were poor, and their proposed model instead generated the two modalities separately.
%using separately trained network models.
The U-Net decoder from Grad-TTS \cite{popov2021grad}, used for audio synthesis in Diff-TTSG \cite{mehta2023diff}, is not a good fit for a unified multimodal decoder either.
A key issue is its use of 2D convolutions, which essentially assume translation invariance not only in time $1{:}T$, but also along the concatenated frequency-and-joint axis (the elements of the vectors $[\boldsymbol{a}^\intercal,\boldsymbol{m}^\intercal]^\intercal$).
This assumption is false due to (1) the presence of concatenation discontinuities and (2) the fact that the joint dimension for motion lacks meaningful translation equivariance
%This assumption is inappropriate both due to the presence of concatenation discontinuities and because the joint dimension for motion does not have meaningful translation equivariance
(see the example features in \cref{fig:architectural overview compact}).
Moreover, the downsampling in the \cite{popov2021grad} U-Net implementation assumes that channel dimensions are divisible by 4, whereas our joint vectors $[\boldsymbol{a}^\intercal,\boldsymbol{m}^\intercal]^\intercal$ had 125 dimensions. 

%Although methods such as \cite{mehta2023diff} produce high-quality motion, they exhibit an information bottleneck. This is because they employ a two-pathway approach where the motion decoder can only access the mean audio, whilst the audio component lacks any interaction with motion.
%Furthermore, these methods do not consider motion when they learn to align input and output sequences; instead, their alignment process is exclusively considers how the input symbols map onto the audio modality (acoustics).
%To alleviate this information bottleneck, we integrated the two pathways and implemented a decoder capable of facilitating interaction between these two representations.
%However, an architecture like the decoder in \cite{popov2021grad} (and also used in \cite{mehta2023diff}) is not a good fit for unified multimodal decoder.
%struggled to capture the desired similarity.
%This was attributed to its use of 2D convolutions, which inherently assume translation invariance in both dimensions including the concatenated frequency-and-joint axis dimension. However, this assumption is not suitable as it disregards the presence of discontinuities at the point of concatenation and the fact that the joint dimension for motion does not have meaningful translation equivariance. Moreover, due to practical constraints, their U-Net architecture failed to accommodate channel dimensions that were not divisible by 4, whereas our joint motion representation had 45 dimensions. 

To enable unified synthesis of both audio and gesture features, the Match-TTSG decoder uses a 1D instead of a 2D U-Net.
After the \texttt{Conv1D} blocks in the decoder U-Net we add Transformers, which may capture longer-range dependencies and still can be made fast; cf.\ \cite{ren2021fastspeech2}.
The resulting self-attention over both modalities should permit the model to better plan gestures based on speech cues and vice versa, for more coherent and contextually relevant output.

%In addition to the ability to sample from the joint multimodal distribution without conditional independence assumptions,
Unifying the audio and motion synthesis pathways also brings several other advantages over Diff-TTSG.
Whereas \cite{wang2021integrated,mehta2023diff} only learnt to align the input and output sequences using audio alone,
%did not consider motion when learning to align the input and output sequences, but only learnt to map input symbols onto audio features,
our proposal makes use of all information (both modalities) for alignment.
This also means that our new encoder directly outputs multimodal features $\boldsymbol{\mu}$,
%The mean squared error (MSE)-based alignment mechanism of \cite{badlani2022one,popov2021grad}, our new encoder directly outputs multimodal features $\boldsymbol{\mu}$
which removes a bottleneck on the information flow in Diff-TTSG, where motion features $\boldsymbol{\mu}_{\boldsymbol{m}}$ were predicted from audio features $\boldsymbol{\mu}_{\boldsymbol{a}}$ by means of an additional Conformer network.
%(not needed in Match-TTSG).
%To alleviate this information bottleneck, we integrated the two pathways and implemented a decoder capable of facilitating interaction between these two representations.
%However, an archi

%We also made some minor architectural innovations apart from the architecture of the unified new decoder.
As a smaller architectural modification, we replaced the conventional relative position encoding in the Diff-TTSG encoder with Rotary Position Embeddings (RoPE) from \cite{su2021roformer}.
These improve memory footprint \cite{wennberg2021case} and generalise better to long utterances \cite{press2022train}.
We found it unnecessary to use position embeddings in the decoder, since positions are implicitly encoded in the encoder output vectors.

The full Match-TTSG model architecture is illustrated in \cref{fig:architectural overview compact}.
This uses the same duration-prediction network as \cite{mehta2023diff}.
%
%To allow for context awareness, we introduced
%and maintaining efficiency, we introduced transformer blocks into our decoder.
%They facilitate complete comprehension of both modalities by allowing the model to capture intricate dependencies and correlations between speech and gestures. Additionally, self-attention over both modalities enables the model to better plan gestures based on speech cues and vice versa, resulting in more coherent and contextually relevant multimodal output. Furthermore, the incorporation of transformers also provides memory computation advantages, adding an additional benefit to their utilisation. To the best of our knowledge, prior to our work, there have been no successful non-autoregressive joint synthesis methods. One notable attempt was a modification to the Glow-TTS architecture in \cite{wang2021integrated}. However, this particular method did not exhibit competitive performance when compared to alternative approaches that were tested, consequently leading to its exclusion from the evaluation altogether.
%
To train the proposed method, we adopt OT-CFM
%framework of conditional flow matching as introduced in
\cite{lipman2023flow},
%as described in \cref{ssec:cfm},
replacing the $\mathcal{L}_{\mathrm{diff}}$ loss term from \cite{popov2021grad}.
%In particular, we employ the Optimal Transport Conditional Flow Matching (OT-CFM) formulation.
%Synthesis begins from $\mathcal{N}(\boldsymbol{0},I)$
Our synthesis process initiates from randomly sampled noise $\mathcal{N}(\boldsymbol{0},I)$ and is conditioned on the output $\boldsymbol{\mu}$ of the encoder after upsampling based on the predicted durations.
%This differs slightly from Diff-TTSG, which samples from $\mathcal{N}(\boldsymbol{\mu},I)$ instead.
%upsampled output of the text encoder, denoted as $\boldsymbol{\mu}$
%(which, due to the MSE-based alignment mechanism \cite{badlani2022one,popov2021grad}, is the conditional average of the output features),
%as illustrated in \cref{fig:architectural overview compact}.

%\begin{figure*}[t]
%  \centering
%  \begin{subfigure}{0.48\textwidth}
%    \includegraphics[width=\linewidth]{training.pdf}
%    \caption{Overview of the training procedure.}
%    \label{fig:training overview}
%  \end{subfigure}
%  \hfill
%  \begin{subfigure}{0.48\textwidth}
%    \includegraphics[width=\linewidth]{synthesis.pdf}
%    \caption{Overview of the synthesis procedure.}
%    \label{fig:synthesis overview}
%  \end{subfigure}
%  \caption{Training and synthesis overview}
%  \label{fig:architectural overview}
%\end{figure*}

% We should mention that simply training Grad-TTS on concatenated features did not work, due to its use of 2D convolutions.
% The downsampling fails because the dimensionality is not divisible by two, but more importantly, 2D convolutions assume translation invariance along the concatenated frequency-and-joint axis, which does not hold at all (cf.\ the example data visualisations in the architecture figure).

\section{Experiments}
\label{sec:experiments}
In this section, we present experiments comparing the proposed method
%in comparison to both natural speech and motion capture, and
to the previous state of the art on the task, Diff-TTSG \cite{mehta2023diff}.
See \href{\webpageurl}{\webpageurltext} for examples and code.

Experimental setup and tests closely mirror those used in \cite{mehta2023diff}, which introduced an improved evaluation suite for integrated speech and gesture synthesis, grounded in prior multimodal work \cite{wang2021integrated} and recent developments in gesture evaluation \cite{jonell2020let,kucherenko2021large,yoon2022genea,kucherenko2023genea,kucherenko2023evaluating}.
For lack of space, we can only outline the data, setup, and user studies, with longer descriptions of the experimental methodology available in \cite{mehta2023diff}, but we give details on any deviations from that prior work.
%Detailed descriptions are confined to deviations from that prior work.
All experiments were performed on NVIDIA RTX 3090 GPUs.

\subsection{Data and systems}
\label{ssec:data}
We used the same data (including splits) as Diff-TTSG, namely the Trinity Speech-Gesture Dataset II (TSGD2)\footnote{\customurl{https://trinityspeechgesture.scss.tcd.ie/}} \cite{ferstl2021expressgesture,ferstl2020adversarial}.
The dataset contains 6 hours of multimodal recordings (time-aligned 44.1 kHz audio and 120 fps marker-based motion-capture) of a male native speaker of Hiberno English, discussing various topics with free gestures.
1.5 hours of this data was held out for testing, with the rest used for training.
We likewise used the same text transcriptions, acoustic features, pose representations (matching the frame rate of the acoustics), and HiFi-GAN vocoder\footnote{\customurl{https://github.com/jik876/hifi-gan/}} \cite{kong2020hifi} as \cite{mehta2023diff} followed by a denoising filter \cite{prenger2019waveglow} with strength \texttt{2e-4}.
We also used the same neutral-looking, skinned upper-body avatar to visualise motion, which omits finger motion (due to inaccurate finger mocap) as well as facial expression, gaze, and lip motion (which are not captured by the dataset nor synthesised by the methods considered).

Since Diff-TTSG has demonstrated better results than all prior methods for integrated speech and gesture synthesis \cite{mehta2023diff}, we only need to compare to Diff-TTSG to demonstrate improvements on the state of the art.
We therefore trained a Diff-TTSG baseline on the data, identical in all respects to that in \cite{mehta2023diff} except the use of better text-processing to transform input graphemes into phonemes (specifically Phonemizer\footnote{\customurl{https://github.com/bootphon/phonemizer/}}
%\cite{bernard2021phonemizer}
with the \texttt{espeak-ng} backend).
We labelled our trained baseline system \textbf{DIFF}.

%System training closely followed \cite{mehta2023diff} with the modification of the front end for our experiments.
%We employed Phonemizer\footnote{\url{https://github.com/bootphon/phonemizer}} \cite{bernard2021phonemizer} with the \texttt{espeak-ng} backend to transform input graphemes into phonemes. We trained previous state-of-the-art Diff-TTSG employing phonemizer as the front-end and labelled this condition as \textbf{DIFF}.

We also trained a system based on the proposed approach.
This system used a text encoder and duration predictor with the same hyperparameters as those in \cite{mehta2023diff}, expect we did not need to specify any window size for relative position encodings as we are using RoPE instead.
The decoder (a.k.a.\ \emph{flow-prediction network} in \cref{fig:architectural overview compact}),
%which takes the form of a 
comprised a 1D Transformer U-Net with two temporal downsampling blocks with hidden dimensionality of 256 and 512, respectively, then two mid-layer blocks with dim.\ 512, and finally, two upsampling blocks with hidden dim.\ 512 and 256.
Each block used one Transformer with four heads, attention dim.\ 64, and `snakebeta' activations \cite{lee2023bigvgan} in the feed-forward network.
The dimensionalities were taken from the U-Net in Diff-TTSG, except that the middle layers were increased from 256 to 512 as our single U-Net is modelling two modalities instead of one.
%inner width of the inner was increased from 256 to 512 for enhanced representational capacity.
%The feed-forward network of each Transformer block used `snakebeta' activations \cite{lee2023bigvgan}.
The output of the U-Net was linearly projected to 80 mel channels and 45 motion features (skeleton joint rotations).
%using a linear projection layer.
We labelled the trained Match-TTSG system \textbf{MA}.

To isolate the benefits of CFM training from those due to the new architecture, we also trained the architecture of system MA using conventional score matching as \cite{mehta2023diff, popov2021grad}, terming the resulting model \textbf{SM} for ``score matching''.
All three systems were trained for 500k updates on 2 GPUs with batch size 32 and learning rate \texttt{1e-4}.

Trained ODE-based models can use a different number of quantisation steps for the numerical ODE solver during sampling, trading off speed against accurate sampling.
Just as in \cite{mehta2023diff}, DIFF used 50 ODE-solver steps for its acoustic pathway but 500 for motion, as using 50 steps did not synthesise smooth motion.
To demonstrate the effectiveness of our proposed architecture with flow matching, we evaluated the results of using both 50 and 500 steps during synthesising from the new systems, denoting these conditions \textbf{MA},\textbf{SM-50} and \textbf{MA},\textbf{SM-500}, respectively.
Fewer steps (e.g., 25) was found to lead to jerky motion also for MA.
In all cases, the first-order Euler forward ODE solver was used.
The evaluations also included a topline condition,
%\textbf{DIFF} denotes the Diff-TTSG baseline and 
% moved introduction of Diff above
\textbf{NAT}, demonstrating copy-synthesis from natural acoustic features and 3D mocap joint rotations in the test set.

\subsection{Evaluation, results, and discussion}
\label{ssec:results}
We evaluated the proposed method both objectively and subjectively.
The objective metrics, displayed in \cref{tab:objective}, show that the proposed method led to a smaller model (fewer parameters) that also required less memory, compared to Diff-TTSG.
%One reason for this is the consolidation of the two separate diffusion pathways into a single pathway, reducing two decoders to one and also eliminating the Conformer network that Diff-TTSG used to predict mean motion from mean acoustics.
This is mainly due to the consolidation of two diffusion pathways into one, but also eliminating the Conformer network that Diff-TTSG uses to predict mean motion from mean acoustics. %The use of 1D rather than 2D convolutions in the decoder means that many thensors have one fewer dimension, resulting
\cref{tab:objective} also lists the real-time factor (RTF) achieved during multimodal synthesis of the 25 test utterances under different conditions.
Clearly, most time is spent solving the ODE.
DIFF is slower than real time, as are MA-500 and SM-500, but using only 50 steps, our proposed architecture runs about five times faster than real time.
%The main reason Diff-TTS uses many ODE-solver steps (500 steps for motion) is because diffusion models generally need many steps to achieve accurate synthesis.
%It is thus central to see if CFM training does not sacrifice quality when fewer steps are used.
%That said,
Additionally, synthesis speed with the new architecture (regardless of training regime) scales better to long utterances: unimodal audio synthesis of a long, 40-second paragraph using 50 steps had RTF 0.13 with DIFF but accelerated to 0.02 for MA/SM-50; see \href{\webpageurl}{our webpage} for the synthesised output.%

To get an impression of the intelligibility of the synthetic speech audio, we synthesised the standard test set from the LJ Speech database\footnote{\customurl{https://keithito.com/LJ-Speech-Dataset/}} (100 utterances), and then ran Whisper \texttt{medium} ASR on it to compute the word error rate (WER) of the resulting automatic transcriptions compared to the input texts.
Such word error rates correlates well with subjective intelligibility \cite{taylor2021confidence}.
%We used the prompts from the LJ Speech test set rather than our own TSGD2 test-set utterances since LJ is more widely used and contains many more sentences (100 versus only 25 for our test set), which reduces the statistical variance of the WER estimates.
The results, in \cref{tab:objective}, suggest that MA is significantly more intelligible than DIFF (9\% versus 12\% WER).
This trend is clear even though the achieved WER numbers are higher than what is seen running Whisper on natural or vocoded LJ Speech audio for the same utterances (cf.\ \cite{mehta2023overflow}).
Several factors may contribute to this, e.g., TSGD2 using an accent with less training data (Hiberno English vs.\ US English), and less well-enunciated mode of speaking (spontaneous vs.\ read speech).%, and the fact that the TSGD2 speaker likely is not included in the Whisper training data, whereas LJ almost certainly is.
\begin{table}[!t]
\centering
\begin{tabular}{@{}l|cc|ccc@{}}
\toprule 
Condition & Params. & RAM & ODE steps & RTF & WER (\%)\tabularnewline
\midrule
DIFF & 44.7M & 14.5 & 50 \& 500 & 1.94 & 12.42\tabularnewline
\midrule
MA-500 & 30.2M & \hphantom{0}8.8 & 500 & 1.76 & \hphantom{0}8.85\tabularnewline
MA-50 & \textquotedbl & \textquotedbl & \hphantom{0}50 & 0.13 & \hphantom{0}8.97\tabularnewline
\midrule
SM-500 & 30.2M & \hphantom{0}8.8 & 500 & 1.78 & 10.39\tabularnewline
SM-50 & \textquotedbl & \textquotedbl & \hphantom{0}50 & 0.18 & 11.74\tabularnewline
\bottomrule
\end{tabular}
\caption{Objective measures. Columns list condition labels, system parameter counts, minimum GPU RAM needed to train (GiB), number of Euler steps, real-time factor on the test set, and ASR WER.%
%The best number in each column is bold.
%NAT had a WER of X.XX.
}
\label{tab:objective}
\vspace{-\baselineskip}
\end{table}

User studies are the gold standard in evaluating synthesis methods.
%The gold standard in evaluating how synthesis is perceived is subjective user studies.
%synthesis tasks is user studies.
%For multimodal synthesis, comprehensive evaluation requires multiple such studies across different modalities.
Comprehensive evaluation (as in, e.g., \cite{wang2021integrated,mitsui2023uniflg,mehta2023diff}) conducts separate user studies evaluating the realism (naturalness for speech and human-likeness for motion) of each ouput modality, and additionally evaluate how appropriate the modalities are for each other, i.e., how well speech and gestures fit together.
In this work, we use the same evaluation protocol as \cite{mehta2023diff}, including the same questions asked, response options, participant recruitment, attention checks and inclusion criteria, as well as statistical analysis. %, with the only differences being the conditions and stimuli in the evaluation and the number of subjects recruited.
In brief, realism was evaluated on unimodal stimuli (audio only, or videos of the avatar moving without audio) being rated on an integer scale 1 through 5 (5 being best) to yield a mean opinion scores (MOS).
For each of the two studies we recruited 35 participants who scored each condition 15 times, giving 525 total ratings of each condition.

Cross-modal appropriateness evaluations, on the other hand, have been found to be confounded by differences in unimodal realism \cite{kucherenko2021large} and verbal content \cite{jonell2020let}.
To control for this, we performed a five-point integer scale pairwise preference tests of multimodal stimuli with matched vs.\ mismatched modalities (the same protocol as in \cite{mehta2023diff}, in turn based on \cite{yoon2022genea,kucherenko2023evaluating,kucherenko2021large}).
A response with clear preference for a mismatched stimulus was coded as $-$2 and one indicating clear preference for the matched stimulus as 2.
525 responses per condition were gathered from a total of 70 participants (twice the number compared to the unimodal studies, since participants now have to watch two videos for each response they provide).
These were averaged to obtain a \emph{mean appropriateness score} (MAS) \cite{kucherenko2023genea} per condition, where a system that generates motion that is completely unrelated to the audio is expected to score 0 on average.

The results of our three user studies are reported in \cref{tab:subjective}.
In the speech-only study, all pairwise differences were statistically significant except (MA-500, MA-50) and (DIFF, SM-50); for the video-only study, only the pairs (MA-500, MA-50) and (DIFF, SM-500) were not significantly different.
%For motion human-likeness, the SM-50 condition was excluded from the evaluation since it was so low quality that including it would have risked reducing the resolution for other differences in the study due to range-compression bias; cf.\ \cite{cooper2023investigating}. %so jerky and low-quality that it was not useful to evaluate, since its presence might reduce the resolution for other differences in the study due to range-compression bias; cf.\ \cite{cooper2023investigating}. SM-50 was consequently also excluded from the multimodal appropriateness evaluation discussed further below. However, we provide example videos of SM-50 on \href{\webpageurl}{our project webpage}.
The SM-50 condition was excluded from any evaluations containing motion due to its low motion quality, but we provide example videos of it on \href{\webpageurl}{our project webpage}.

The unimodal evaluations exhibited similar trends for both audio and video.
First, the two MA conditions consistently improved on DIFF in both evaluations, meaning that the proposed approach led to significantly more realistic synthesis.
Second, the two MA conditions achieved very similar numerical scores in both evaluations.
%in contrast to when the same architecture was trained by score matching, where SM-500 scored much better than SM-50.
When instead training same architecture using score matching, SM-500 scored much better than SM-50.
This indicates that OT-CFM training not only improves synthesis quality, but also retains quality with much fewer steps.
%These findings indicate that OT-CFM training not only improves synthesis quality, but that an accurate solution of the associated ODE is possible in much fewer steps than with score-matching.
%suggests that the new.
For audio, DIFF (using 50 steps) and SM-50 performed similarly (no statistically significant difference); for motion, DIFF (using 500 steps) and SM-500 performed similarly.
This suggests that the new architecture, whilst more compact, is similarly effective as the individual synthesis pathways in Diff-TTSG.

The multimodal appropriateness evaluation found all pairwise differences between systems to be statistically significant, except those between systems in the set \{MA-500, MA-50, SM-500\}.
In particular, the proposed system demonstrated significantly better appropriateness than DIFF at both ODE-solver step counts.
Furthermore, these results indicate that the unified architecture, although not a significant improvement
%on naturalness/human-likeneness
in the unimodal evaluations, contributed significantly to the improved appropriateness, as all systems using the new architecture were significantly better than DIFF.
This is reasonable, given that the proposed architecture samples from the joint distribution of the two modalities whereas Diff-TTSG samples from two conditionally independent marginals.
In practice, the numerical difference in MAS between SM-500 and the two MA systems suggests that the training regimen also may play a part in the overall appropriateness-score improvement, though our statistical test does not resolve the two with the current size of the subjective evaluation.

Taken together, we can conclude that flow matching (OT-CFM) training enables better synthesis realism in much fewer synthesis steps, whilst the new architecture is memory and parameter efficient, faster on long utterances, and improves cross-modal appropriateness, important for effective communication in a multimodal agent.
\begin{table}[!t]
\centering
\begin{tabular}{@{}l|cc|c@{}}
\toprule 
Condition & Speech MOS & Motion MOS & MAS (multimodal)\tabularnewline
\midrule
NAT & 4.35$\pm$0.07 & 4.01$\pm$0.10 & 1.16$\pm$0.10\tabularnewline
DIFF & 3.27$\pm$0.10 & 3.11$\pm$0.10 & 0.31$\pm$0.09\tabularnewline
\midrule
MA-500 & 3.70$\pm$0.08 & 3.44$\pm$0.10 & 0.53$\pm$0.09\tabularnewline
MA-50 & 3.67$\pm$0.09 & 3.43$\pm$0.09 & 0.52$\pm$0.09\tabularnewline
\midrule
SM-500 & 3.38$\pm$0.09 & 3.13$\pm$0.09 & 0.43$\pm$0.09\tabularnewline
SM-50 & 3.22$\pm$0.09 & N/A & N/A\tabularnewline
\bottomrule
\end{tabular}
\caption{Results of subjective evaluations. All mean opinion scores (MOS, 1 to 5, unimodal) and mean appropriateness scores (MAS, $-$2 to 2, multimodal) come with 95\%-confidence intervals.}
\label{tab:subjective}
\vspace{-\baselineskip}
\end{table}

\section{Conclusions and future work}
\label{sec:conclusion}
We have introduced Match-TTSG, a new approach to multimodal synthesis of speech acoustics and 3D co-speech gesture motion.
The key changes are a new, unified decoder network for joint synthesis of both modalities, and the use of optimal-transport conditional flow matching for model training.
The unified architecture allows sampling from the joint distribution of both modalities instead of (as previously) two conditionally independent marginals, improving cross-modal appropriateness,
The new training methodology allows high output quality using much fewer ODE-solver steps, i.e., faster synthesis.
The new method was found to improve the multimodal synthesis of acoustics and 3D motion on all objective and subjective metrics, compared to the previous state of the art, Diff-TTSG.

For future work, it would be interesting to incorporate self-supervised representations learnt from larger datasets, similar to \cite{wang2023use,deichler2023diffusion}, and to add even more modalities to the synthesis.

% Gustav Eje Henter 2023-08-08: I changed this to a modern IEEEtran.bst
% This gives identical results but automatically abbreviates author names
% -------------------------------------------------------------------------
\bibliographystyle{IEEEtran}
\bibliography{refs_abbrev}

\begin{thebibliography}{10}
\providecommand{\url}[1]{#1}
\csname url@samestyle\endcsname
\providecommand{\newblock}{\relax}
\providecommand{\bibinfo}[2]{#2}
\providecommand{\BIBentrySTDinterwordspacing}{\spaceskip=0pt\relax}
\providecommand{\BIBentryALTinterwordstretchfactor}{4}
\providecommand{\BIBentryALTinterwordspacing}{\spaceskip=\fontdimen2\font plus
\BIBentryALTinterwordstretchfactor\fontdimen3\font minus
  \fontdimen4\font\relax}
\providecommand{\BIBforeignlanguage}[2]{{%
\expandafter\ifx\csname l@#1\endcsname\relax
\typeout{** WARNING: IEEEtran.bst: No hyphenation pattern has been}%
\typeout{** loaded for the language `#1'. Using the pattern for}%
\typeout{** the default language instead.}%
\else
\language=\csname l@#1\endcsname
\fi
#2}}
\providecommand{\BIBdecl}{\relax}
\BIBdecl

\bibitem{mcneill2008gesture}
D.~McNeill, \emph{Gesture and Thought}.\hskip 1em plus 0.5em minus 0.4em\relax
  U. Chicago Press, 2008.

\bibitem{kendon1988gestures}
A.~Kendon, ``How gestures can become like words,'' in \emph{Cross-Cultural
  Perspectives in Nonverbal Communication}, 1988.

\bibitem{nyatsanga2023comprehensive}
S.~Nyatsanga, T.~Kucherenko, C.~Ahuja, G.~E. Henter, and M.~Neff, ``A
  comprehensive review of data-driven co-speech gesture generation,''
  \emph{Comput. Graph. Forum}, 2023.

\bibitem{wagner2014gesture}
P.~Wagner, Z.~Malisz, and S.~Kopp, ``Gesture and speech in interaction: An
  overview,'' \emph{Speech Commun.}, vol.~57, 2014.

\bibitem{shen2018natural}
J.~Shen, R.~Pang, R.~J. Weiss, M.~Schuster, N.~Jaitly, Z.~Yang, Z.~Chen
  \emph{et~al.}, ``Natural {TTS} synthesis by conditioning {W}ave{N}et on mel
  spectrogram predictions,'' in \emph{Proc. ICASSP}, 2018.

\bibitem{mehta2023overflow}
S.~Mehta, A.~Kirkland, H.~Lameris, J.~Beskow, {\'E}.~Sz{\'e}kely, and G.~E.
  Henter, ``{O}ver{F}low: Putting flows on top of neural transducers for better
  {TTS},'' in \emph{Proc. Interspeech}, 2023.

\bibitem{szekely2020breathing}
{\'E}.~Sz{\'e}kely, G.~E. Henter, J.~Beskow, and J.~Gustafson, ``Breathing and
  speech planning in spontaneous speech synthesis,'' in \emph{Proc. ICASSP},
  2020, pp. 7649--7653.

\bibitem{alexanderson2023listen}
S.~Alexanderson, R.~Nagy, J.~Beskow, and G.~E. Henter, ``Listen, denoise,
  action! {A}udio-driven motion synthesis with diffusion models,'' \emph{ACM
  ToG}, vol.~42, no.~4, 2023, article 44.

\bibitem{ao2023gesturediffuclip}
T.~Ao, Z.~Zhang, and L.~Liu, ``{G}esture{D}iffu{CLIP}: Gesture diffusion model
  with {CLIP} latents,'' \emph{ACM ToG}, vol.~42, no.~4, 2023, article 42.

\bibitem{yoon2022genea}
Y.~Yoon, P.~Wolfert, T.~Kucherenko, C.~Viegas, T.~Nikolov, M.~Tsakov, and G.~E.
  Henter, ``{T}he {GENEA} {C}hallenge 2022: {A} large evaluation of data-driven
  co-speech gesture generation,'' in \emph{Proc. ICMI}, 2022, pp. 736--747.

\bibitem{mehta2023diff}
S.~Mehta, S.~Wang, S.~Alexanderson, J.~Beskow, {\'E}.~Sz{\'e}kely, and G.~E.
  Henter, ``{D}iff-{TTSG}: Denoising probabilistic integrated speech and
  gesture synthesis,'' in \emph{Proc. SSW}, 2023.

\bibitem{kucherenko2023genea}
T.~Kucherenko, R.~Nagy, Y.~Yoon, J.~Woo, T.~Nikolov, M.~Tsakov, and G.~E.
  Henter, ``The {GENEA} {C}hallenge 2023: {A} large-scale evaluation of gesture
  generation models in monadic and dyadic settings,'' in \emph{Proc. ICMI},
  2023.

\bibitem{lipman2023flow}
Y.~Lipman, R.~T.~Q. Chen, H.~Ben-Hamu \emph{et~al.}, ``Flow matching for
  generative modeling,'' in \emph{Proc. ICLR}, 2023.

\bibitem{song2021score}
Y.~Song, J.~Sohl-Dickstein, D.~P. Kingma, A.~Kumar, S.~Ermon, and B.~Poole,
  ``Score-based generative modeling through stochastic differential
  equations,'' in \emph{Proc. ICLR}, 2021.

\bibitem{popov2021grad}
V.~Popov, I.~Vovk, V.~Gogoryan, T.~Sadekova, and M.~Kudinov, ``Grad-{TTS}: A
  diffusion probabilistic model for text-to-speech,'' in \emph{Proc. ICML},
  2021, pp. 8599--8608.

\bibitem{kim2021vits}
J.~Kim, J.~Kong, and J.~Son, ``{VITS}: Conditional variational autoencoder with
  adversarial learning for end-to-end text-to-speech,'' in \emph{Proc. ICML},
  2021, pp. 5530--5540.

\bibitem{alexanderson2020style}
S.~Alexanderson, G.~E. Henter, T.~Kucherenko, and J.~Beskow,
  ``Style-controllable speech-driven gesture synthesis using normalising
  flows,'' \emph{Comput. Graph. Forum}, vol.~39, no.~2, 2020.

\bibitem{yu2019durian}
C.~Yu, H.~Lu, N.~Hu, M.~Yu, C.~Weng \emph{et~al.}, ``{D}ur{IAN}: Duration
  informed attention network for multimodal synthesis,'' \emph{arXiv preprint
  arXiv:1909.01700}, 2019.

\bibitem{mitsui2023uniflg}
K.~Mitsui, Y.~Hono, and K.~Sawada, ``{U}ni{FLG}: Unified facial landmark
  generator from text or speech,'' in \emph{Proc. Interspeech}, 2023, pp.
  5501--5505.

\bibitem{salem2010towards}
M.~Salem, S.~Kopp, I.~Wachsmuth, and F.~Joublin, ``Towards an integrated model
  of speech and gesture production for multi-modal robot behavior,'' in
  \emph{Proc. RO-MAN}, 2010, pp. 614--619.

\bibitem{alexanderson2020generating}
S.~Alexanderson, {\'E}.~Sz{\'e}kely, G.~E. Henter, T.~Kucherenko, and
  J.~Beskow, ``Generating coherent spontaneous speech and gesture from text,''
  in \emph{Proc. IVA}, 2020, pp. 1--3.

\bibitem{wang2021integrated}
S.~Wang, S.~Alexanderson, J.~Gustafson, J.~Beskow, G.~E. Henter, and
  {\'E}.~Sz\'{e}kely, ``Integrated speech and gesture synthesis,'' in
  \emph{Proc. ICMI}, 2021, pp. 177--185.

\bibitem{kim2020glow}
J.~Kim, S.~Kim, J.~Kong, and S.~Yoon, ``Glow-{TTS}: A generative flow for
  text-to-speech via monotonic alignment search,'' in \emph{Proc. NeurIPS},
  2020, pp. 8067--8077.

\bibitem{chen2018neural}
R.~T.~Q. Chen, Y.~Rubanova, J.~Bettencourt \emph{et~al.}, ``Neural ordinary
  differential equations,'' in \emph{Proc. NeurIPS}, 2018.

\bibitem{liu2023flow}
X.~Liu, C.~Gong, and Q.~Liu, ``Flow straight and fast: Learning to generate and
  transfer data with rectified flow,'' in \emph{Proc. ICLR}, 2023.

\bibitem{le2023voicebox}
M.~Le, A.~Vyas, B.~Shi, B.~Karrer, L.~Sari, R.~Moritz \emph{et~al.},
  ``Voicebox: Text-guided multilingual universal speech generation at scale,''
  \emph{arXiv preprint arXiv:2306.15687}, 2023.

\bibitem{hu2023motion}
V.~T. Hu, W.~Yin, P.~Ma, Y.~Chen, B.~Fernando, Y.~M. Asano, E.~Gavves
  \emph{et~al.}, ``Motion flow matching for human motion synthesis and
  editing,'' \emph{arXiv preprint arXiv:2312.08895}, 2023.

\bibitem{ren2021fastspeech2}
Y.~Ren, C.~Hu, X.~Tan \emph{et~al.}, ``{F}ast{S}peech 2: Fast and high-quality
  end-to-end text to speech,'' in \emph{Proc. ICLR}, 2021.

\bibitem{su2021roformer}
J.~Su, Y.~Lu, S.~Pan, A.~Murtadha, B.~Wen, and Y.~Liu, ``{R}o{F}ormer: Enhanced
  {T}ransformer with rotary position embedding,'' \emph{arXiv preprint
  arXiv:2104.09864}, 2021.

\bibitem{wennberg2021case}
U.~Wennberg and G.~E. Henter, ``The case for translation-invariant
  self-attention in {T}ransformer-based language models,'' in \emph{Proc.
  ACL-IJCNLP Vol. 2}, 2021, pp. 130--140.

\bibitem{press2022train}
O.~Press, N.~A. Smith, and M.~Lewis, ``Train short, test long: Attention with
  linear biases enables input length extrapolation,'' in \emph{Proc. ICLR},
  2022.

\bibitem{jonell2020let}
P.~Jonell, T.~Kucherenko, G.~E. Henter, and J.~Beskow, ``Let's face it:
  Probabilistic multi-modal interlocutor-aware generation of facial gestures in
  dyadic settings,'' in \emph{Proc. IVA}, 2020.

\bibitem{kucherenko2021large}
T.~Kucherenko, P.~Jonell, Y.~Yoon, P.~Wolfert, and G.~E. Henter, ``A large,
  crowdsourced evaluation of gesture generation systems on common data: The
  {GENEA} {C}hallenge 2020,'' in \emph{Proc. IUI}, 2021, pp. 11--21.

\bibitem{kucherenko2023evaluating}
T.~Kucherenko, P.~Wolfert, Y.~Yoon, C.~Viegas, T.~Nikolov, M.~Tsakov, and G.~E.
  Henter, ``Evaluating gesture-generation in a large-scale open challenge: The
  {GENEA} {C}hallenge 2022,'' \emph{arXiv preprint arXiv:2303.08737}, 2023.

\bibitem{ferstl2021expressgesture}
Y.~Ferstl, M.~Neff, and R.~McDonnell, ``{E}xpress{G}esture: Expressive gesture
  generation from speech through database matching,'' \emph{Comput. Animat.
  Virt. W.}, p. e2016, 2021.

\bibitem{ferstl2020adversarial}
------, ``Adversarial gesture generation with realistic gesture phasing,''
  \emph{Comput. Graph.}, vol.~89, pp. 117--130, 2020.

\bibitem{kong2020hifi}
J.~Kong, J.~Kim, and J.~Bae, ``{H}i{F}i-{GAN}: Generative adversarial networks
  for efficient and high fidelity speech synthesis,'' in \emph{Proc. NeurIPS},
  2020, pp. 17\,022--17\,033.

\bibitem{prenger2019waveglow}
R.~Prenger, R.~Valle \emph{et~al.}, ``{W}ave{G}low: A flow-based generative
  network for speech synthesis,'' in \emph{Proc. ICASSP}, 2019.

\bibitem{lee2023bigvgan}
S.-g. Lee, W.~Ping, B.~Ginsburg, B.~Catanzaro, and S.~Yoon, ``{B}ig{VGAN}: A
  universal neural vocoder with large-scale training,'' in \emph{Proc. ICLR},
  2023.

\bibitem{taylor2021confidence}
J.~Taylor and K.~Richmond, ``Confidence intervals for {ASR}-based {TTS}
  evaluation,'' in \emph{Proc. Interspeech}, 2021.

\bibitem{wang2023use}
S.~Wang, G.~E. Henter, J.~Gustafson, and {\'E}.~Sz{\'e}kely, ``On the use of
  self-supervised speech representations in spontaneous speech synthesis,'' in
  \emph{Proc. SSW}, 2023.

\bibitem{deichler2023diffusion}
A.~Deichler, S.~Mehta, S.~Alexanderson, and J.~Beskow, ``Diffusion-based
  co-speech gesture generation using joint text and audio representation,'' in
  \emph{Proc. ICMI}, 2023.

\end{thebibliography}

\end{document}